\documentclass[aps,prc,reprint,superscriptaddress,showpacs,amsmath,amssymb,longbibliography,lengthcheck]{revtex4-1}

\usepackage[]{graphicx}
\usepackage{dcolumn}
\usepackage{txfonts}
\begin{document}

 \title{The contrasting fission potential-energy structures of 
actinides and mercury isotopes} 

\author{Takatoshi Ichikawa}%
\affiliation{Yukawa Institute for Theoretical Physics, Kyoto University,
Kyoto 606-8502, Japan}
\author{Akira Iwamoto}
\affiliation{Advanced Science Research Center, Japan
Atomic Energy Agency, Tokai-mura, Naka-gun, Ibaraki 319-1195, Japan}
\author{Peter M\"oller}
\affiliation{Theoretical Division, Los Alamos National Laboratory, Los
Alamos, New Mexico 87545, USA}
\author{Arnold J.\ Sierk}
\affiliation{Theoretical Division, Los Alamos National Laboratory, Los
Alamos, New Mexico 87545, USA}
\date{\today}

\begin{abstract}
{\bf Background:}
Fission-fragment mass distributions are asymmetric in fission of
 typical actinide nuclei for nucleon number $A$ in the range $228
 \lnsim A \lnsim 258$ and proton number $Z$ in the range $90\lnsim Z
 \lnsim 100$. For somewhat lighter systems it has been observed that
 fission mass distributions are usually symmetric. However, a recent
 experiment showed that fission of $^{180}$Hg following electron
 capture on $^{180}$Tl is asymmetric.  
{\bf Purpose:} We calculate potential-energy surfaces for a
 typical actinide nucleus and for 12 even isotopes in the range
 $^{178}$Hg--$^{200}$Hg, to investigate the similarities and
 differences of actinide compared to mercury potential surfaces
 and to what extent fission-fragment properties,
 in particular shell structure,
 relate to the structure of the static potential-energy surfaces.
 {\bf Methods:} Potential-energy surfaces are calculated
in the macroscopic-microscopic approach as functions of five
shape coordinates for more than five million shapes.
The structure of the surfaces are investigated by use of an immersion
technique. {\bf Results:} We determine properties of
minima, saddle points, valleys, and ridges between valleys in
the 5D shape-coordinate space. Along the mercury isotope chain
the barrier heights and the ridge heights and persistence with
elongation vary significantly and show no obvious connection
to possible fragment shell structure, in contrast to the actinide
region, where there is a deep
asymmetric valley extending from the saddle point to scission.
{\bf Conclusions:}  The mechanism of asymmetric fission
must be very different in the lighter proton-rich mercury isotopes compared to
the actinide region and is apparently unrelated to fragment shell structure.
Isotopes lighter than $^{192}$Hg have the saddle point blocked from
a deep symmetric valley by a significant ridge. The ridge vanishes
for the heavier Hg isotopes, for which we would expect a qualitatively
different asymmetry of the fragments.

\end{abstract}

\maketitle

\section{Introduction}
The evolution of a nucleus from a single ground-state shape into two
separated fragments in nuclear fission has, since its discovery
\cite{hahn39:a}, been described in terms of potential-energy surfaces
that are functions of suitable shape coordinates
\cite{meitner39:a,bohr39:a}.  Originally the potential energy was
modeled in terms of a macroscopic liquid-drop model
\cite{meitner39:a,bohr39:a,frankel47:a,hill53:a}.  Subsequently it
became clear that the liquid-drop model cannot explain many features of
fission such as the fission-fragment mass yields, fission-barrier
structure, and actinide fission half-lives
\cite{hill53:a,swiatecki55:a,strutinsky67:a,strutinsky67:b,strutinsky68:a,moller70:a,moller01:a,moller09:a},
because microscopic shell effects significantly perturb the energy
surface given by the liquid-drop model. Although the energy release in
fission, that is the potential-energy change between the ground state of
a single system and well-separated fragments, is more than 200 MeV,
microscopic effects in the narrow range of zero to ten MeV can affect
half-lives by more than ten orders of magnitude and change
fission-fragment mass yields from symmetric to significantly asymmetric.

Experimental observations are that fission-fragment mass distributions
are asymmetric in low-energy fission of typical actinide nuclei for
nucleon number $A$ in the range $228 \lnsim A \lnsim 258$ and proton
number $Z$ in the range $90\lnsim Z \lnsim 100$.  In those nuclei, it
has been established that the heavy-mass peak in the yield distribution
is close to $A=140$, independently of fissioning system, see for example
\cite{wilets64:a,vandenbosch73:a}.  This was thought to originate from
the strong spherical shell effects present in fragments near the doubly
magic nucleus $^{132}_{\phantom{0}50}$Sn$_{82}$, although we now know
that an analysis of high-dimensional potential-energy surfaces 
\cite{moller01:a,moller09:a}, coupled
with a dynamical description is required to robustly establish this
connection \cite{randrup11:a}. In particular, we
now know that ``fragment-shell'' arguments or saddle-point properties
cannot by themselves reliably predict the degree of asymmetry; rather,
the character of the entire potential-energy surface between the
ground-state and separated fragments must be considered
\cite{andreyev10:a,randrup11:a}.

A large-scale experiment studying fission of nuclei in the region
$205\leq A \leq 234$ showed that a transition to symmetric fission
occurred just below the actinide region and that fission remained
symmetric at least down to proton number $Z=85$ and nucleon number
$A=205$.  The dividing line between asymmetric and symmetric fission was
found to approximately follow constant nucleon number, $A=226$
\cite{schmidt01:a}.  The position of this transition line
was predicted to within about 2 neutrons in a 
simple static calculation in 1972 \cite{moller72:a}. 
For slightly lighter systems
\cite{itkis90:a,itkis91:a} near $Z=82$ and $A=200$, a hint of asymmetric
fission was observed for energies up to about 10 MeV above the
saddle-point energy.  Itkis referred to this as ``asymmetry of symmetric
fission'' \cite{itkis90:a}, so it is unclear whether or not he viewed
his results as a clear indication of the onset of a new region of
asymmetric fission.  Despite this intriguing result, it has often been
assumed that fission mass distributions for systems below the actinide
region would be symmetric because, based on the proton and neutron
numbers of possible compound systems, division into fragments with $Z$
and $N$ sufficiently close to $^{132}$Sn (or to much lighter doubly
magic nuclides) so as to exhibit strong shell effects appeared not
possible for almost all compound systems below $A\approx 200$.
Surprisingly, a recent experiment showed \cite{andreyev10:a} that
fission of $^{180}$Hg following electron capture by $^{180}$Tl is
asymmetric.
   
It was earlier argued that the asymmetric fission of $^{180}$Hg was a
new type of asymmetric fission with its origins in the local structure
of the fission potential-energy surface near the fission saddle point
\cite{andreyev10:a}.  Moreover, it was argued that these observations
showed that consideration of ``fragment shells'' does not offer a
general method of predicting or explaining asymmetry in fission.

To illustrate the contrasting origins of asymmetric fission in the Hg
and actinide regions, we calculate and analyze the structure of
five-dimensional fission potential-energy surfaces for even Hg isotopes
in the range $ 178 \leq A \leq 200 $ and compare them to a typical
actinide potential-energy surface, namely that of $^{236}$U.

\begin{figure}[t]
\includegraphics[keepaspectratio,width=\linewidth]{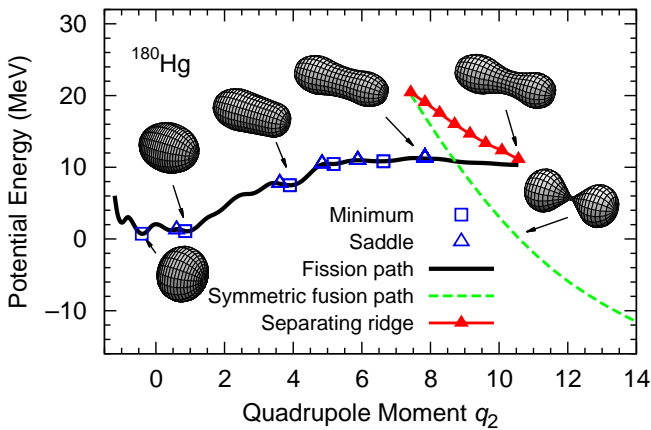}
 \caption{
(Color online) Potential-energy curves, minima, saddles, and ridges for
$^{180}$Hg versus $q_{\rm 2}$ from oblate shapes to very deformed
configurations.  The solid line denotes the optimum fission path leading
to a mass-asymmetric split.  The gray (green) dashed line denotes a
symmetric valley in the potential-energy surface, corresponding to a
compact fusion valley with zero-radius neck shapes along the entire
valley.  The solid (red) line with superimposed triangles is the ridge
separating those two channels.}  \label{Hg180-1d}
\end{figure}

\begin{figure}[t]
\includegraphics[keepaspectratio,width=\linewidth]{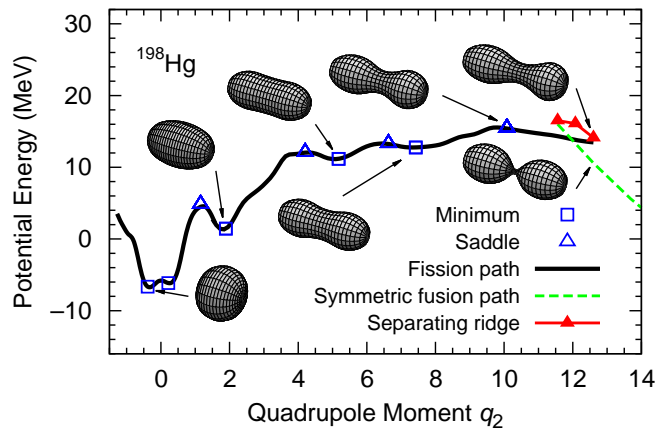}\\%
\caption{(Color online) Potential-energy curves, minima, saddles and
ridges for $^{198}$Hg versus $q_{\rm 2}$ from oblate shapes to very
deformed configurations.  The dashed line is a symmetric valley
corresponding to a fusion valley with deformed shapes connected by a
conicoidal neck region.  The symbols are the same as
Fig.~\ref{Hg180-1d}.}  \label{Hg198-1d}
\end{figure}

\section{Model}
The potential energy is calculated in the FRLDM
\cite{moller95:b,moller09:a}, with the 2002 parameter set for
the macroscopic model \cite{moller04:a}. 
We use two shape parameterizations. For more elongated
shapes somewhat beyond the ground state 
we use the three-quadratic-surface (3QS)
parametrization~\cite{nix68:a,nix69:a} to describe nuclear shapes in a
five-dimensional deformation space. The shape degrees of freedom are a
quadrupole-moment parameter $q_2$, a neck-related parameter $\eta$,
heavier- and lighter-fragment deformation parameters $\epsilon_{\rm H}$ and
$\epsilon_{\rm L}$, and a mass-asymmetry parameter $\alpha_{\rm g}$.
The parameter $\eta$ is related to the curvature of the middle body.
The parameter $q_2$ is the dimensionless quadrupole moment in units of
$3ZR_{0}^2/4\pi(e^2b)$, where $Z$ is the proton number and $R_0$ is the
nuclear radius.  The parameter $\epsilon$ is the Nilsson
perturbed-spheroid parameter.  The mass-asymmetry parameter is
$\alpha_{\rm g}=(M_{\rm H} - M_{\rm L})/(M_{\rm H} + M_{\rm L})$, where
$M_{\rm H}$ and $M_{\rm L}$ are the masses of the heavier and lighter nascent
fragments, respectively.  For finite neck radii these masses are defined
as discussed in \cite{moller01:a}.  The microscopic single-particle
potential is calculated by folding a Yukawa function over the shape of a
``sharp-surface generating volume'' \cite{bolsterli72:a}.

We calculate the adiabatic potential-energy surfaces in this
five-dimensional deformation space for the 12 even isotopes in the range
$^{178-200}$Hg and for $^{236}$U and analyze their structure using the
immersion method \cite{moller09:a}.  The potential energies are
determined at $41\times15\times15\times15\times35$ grid points for $q_2
\times\eta\times\epsilon_{\rm H}\times\epsilon_{\rm L}\times\alpha_{\rm
g}$.  For $q_2$ and $\eta$ we use similar, and for fragment deformations
and asymmetry $\alpha_{\rm g}$, exactly the same points as in
Ref.~\cite{moller09:a}.  We take into account the shape-dependent Wigner
and $A^{0}$ terms in our calculations~\cite{moller09:a}.  

Near the
ground states where $q_2 \leq 0.5$, we also perform complementary
constrained-multipole ($\beta_2$) calculations, which better describe
compact shapes for small deformations \cite{moller95:b}.  We identify
the minima and potential valleys under the condition that their depths
are deeper than 0.05 and 0.2 MeV, respectively.  In our static studies
we can make realistic determinations of major features in the
potential-energy surfaces, such as minima, saddles, valleys, and ridges
between valleys, because in our model we 1) calculate the energy in
millions of grid points for the five most essential shape degrees of
freedom and 2) use an immersion method to extract structure features
\cite{moller09:a}. In contrast, in self-consistent methods in which {\it
constraints} are imposed, the inferred saddle points and ridges may be
overestimated by amounts that can be quite large. Moreover, the
magnitude of this overestimation is impossible to determine, see
Ref. \cite{moller09:a} for a detailed discussion.

\section{Calculated structure of  potential-energy surfaces}
In the early days of theoretical fission studies based on the
macroscopic-microscopic method, most or all investigations calculated
the fission potential-energy surface in terms of only two independent
shape variables, for example variables related to elongation and neck
radius \cite{strutinsky67:a,nilsson69:a} or elongation and fragment mass
asymmetries \cite{moller70:a}.  Complete results from such calculations
could be faithfully displayed in terms of two-dimensional contour
diagrams. In contrast, it is impossible to show all essential features
of five-dimensional potential surfaces by reducing them to
two-dimensional contour plots.  To identify major features of the
5D spaces we start by locating all minima, saddles, ridges and valleys
by use of the immersion technique; for details see
Ref.~\cite{moller09:a}. We then show features identified to be of
special interest in one-dimensional plots versus quadrupole moment. For
example, we show the energies along specific one-dimensional paths,
such as valleys and ridges,
embedded in the full 5D space and relevant
minima and saddles. To more clearly visualize the substantial
differences of asymmetric fission in the neutron-deficient Hg region and
actinide region we will also plot 2D surfaces embedded in the full 5D
deformation space.

\begin{figure}[t]
\includegraphics[keepaspectratio,width=\linewidth]{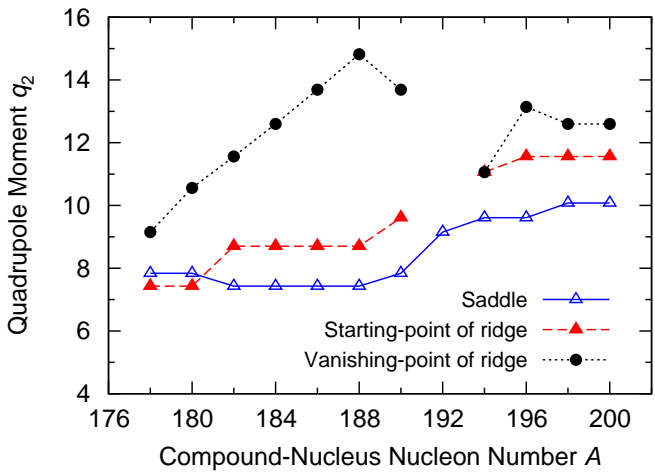}\\%
\caption{(Color online) Saddle and ridge locations for a range of Hg
isotopes. An extended ridge is only present for isotopes in the interval
$180\leq A \leq190$.  For $A=192$ we could not clearly interpret the
ridge features so data are omitted for this isotope.  } \label{q2dep}
\end{figure}
\begin{figure}[t]
\includegraphics[keepaspectratio,width=\linewidth]{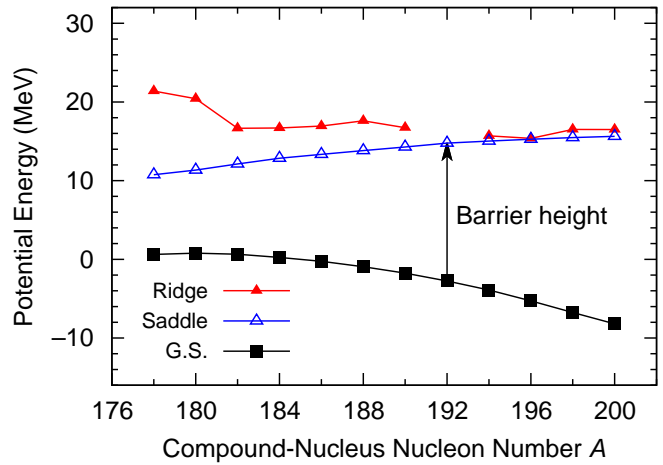}\\%
\caption{(Color online) Ground-state energy, saddle energy
and maximum ridge height, all with respect to the spherical macroscopic
energy.  The difference between the saddle energy and the ground-state
energy is the barrier height, as indicated with an arrow for $^{192}$Hg.
It is only for isotopes in the interval $178\leq A \leq190$ that a ridge
rises above the saddle.}  \label{sadhgt}
\end{figure}

\begin{figure}[t]
\includegraphics[keepaspectratio,width=\linewidth]{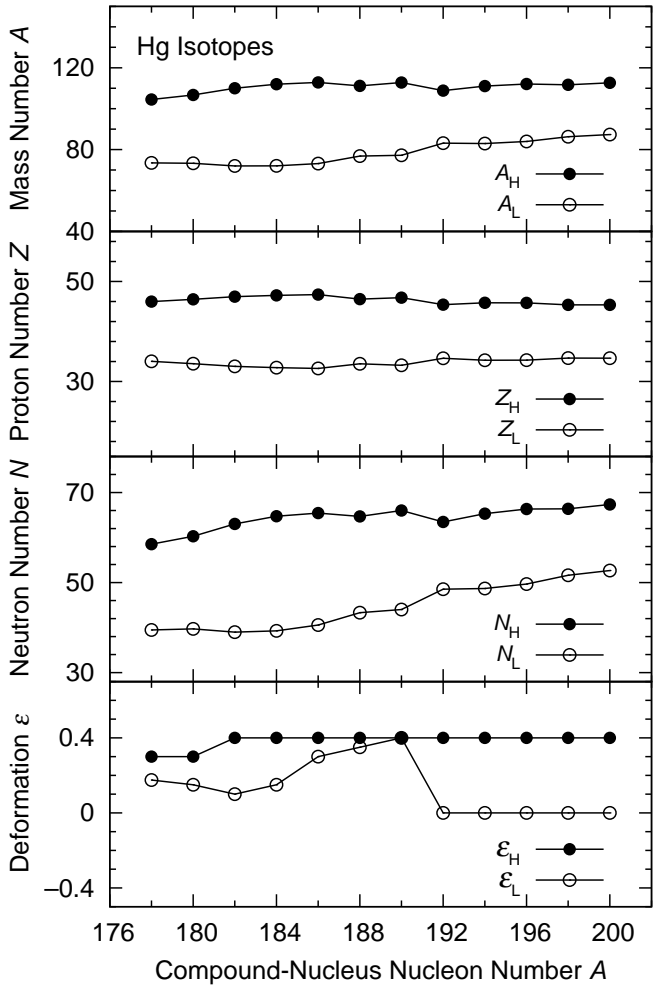}\\%
\caption{Mass number (top), proton number (second), neutron number
(third), and deformation (bottom) of the heavy and light nascent fission
fragments at the vanishing point of the separating ridge determined from
wave-function densities in the two fragments by methods described in
\cite{ichikawa09:a}.  } \label{wavedens}
\end{figure}
\begin{figure}[t]
\includegraphics[keepaspectratio,width=\linewidth]{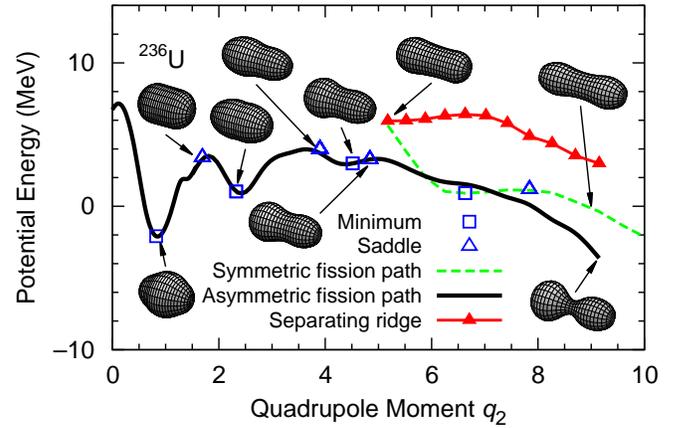}\\%
\caption{(Color online) Potential-energy curves, minima, saddles, and
ridges for $^{236}$U from a spherical shape to very deformed
configurations.  Here the symmetric valley is well separated from the
asymmetric valley by a ridge that is about 5 MeV high along the entire
deformation range between the saddle and the asymmetric scission
configuration.}  \label{U236-1d}
\end{figure}

Figures \ref{Hg180-1d} and \ref{Hg198-1d} show calculated ``optimal''
one-dimensional potential-energy curves or ``fission barriers'',
embedded in the five-dimensional space, as functions of $q_2$ (solid
line) for $^{180}$Hg and $^{198}$Hg. In this study, all potential
energies are measured from the spherical macroscopic energy.
Minima and saddle points are
indicated by open squares and triangles, respectively.  Shapes of the
nuclear macroscopic densities at several saddle points and minima are
also displayed.

In both systems the ground-state shapes are slightly oblate.  However,
the density evolutions from the ground state to the fission saddle
points differ substantially.  For $^{180}$Hg, mass asymmetry has
developed already near the local energy minimum at $q_2=4.0$, although
no distinct fragments have yet emerged, cf. Fig.~\ref{Hg180-1d}.
Subsequently the neck develops, while the degree of mass asymmetry is
retained. At the fission saddle point $E_{\rm sad}=11.35$ MeV, and its
shape corresponds to $q_2=7.84$, $\epsilon_{\rm H}= 0.275$,
$\epsilon_{\rm L}= 0.30$, and $\alpha_{\rm g}=0.14$ or equivalently
$A_{\rm H}/A_{\rm L}= 102.6/77.4$.

On the other hand, the shape for $^{198}$Hg remains symmetric up to the
local energy minimum at $q_2=7.5$, although the neck is well developed
there.  Beyond this local minimum, the mass asymmetry of the fissioning
nuclei develops in tandem with neck formation.  At the fission saddle
point $E_{\rm sad}=15.47$ MeV, $q_2=10.08$, $\epsilon_{\rm H}= 0.35$,
$\epsilon_{\rm L}=0.10$, and $\alpha_{\rm g}=0.12$, or equivalently
$A_{\rm H}/A_{\rm L}= 110.9/87.1$.

In the outer saddle region additional valleys appear in the two
potential-energy surfaces. For each of the two systems we show only one
of these valleys, namely the one corresponding to symmetric shapes as
dashed (green) lines. To leave the figures uncluttered we do not show an
asymmetric valley which is also present.  Often these valleys are
referred to as {\it fusion} valleys because along the entire curve the
neck radius is zero. In a more general treatment allowing for a family
of shapes of separated nuclei, the fragments, or equivalently, the two
colliding heavy ions would be separated along this curve until they have
approached sufficiently close that they touch. Separated fragments are
inaccessible in the 3QS parameterization in its current implementation.
Instead these configurations are represented as two spheroidal nascent
fragments connected by a conicoidal neck \cite{nix68:a,nix69:a}.  This
limitation does not affect our study here, since we only follow the
shape evolution until just before zero neck radius (in a more general
treatment, separation) occurs. What we wish to establish here is the
structure of the potential-energy surface from outside the saddle point
to just before separation.  Is it possible to determine if it favors
evolution towards the symmetric valley or the asymmetric valley? And
when is the final fragment asymmetry established? Clearly it will be
frozen in prior to reaching the bottom of any of the valleys, since
zero-neck-radius shapes occur already above the valley floors.  For
$^{180}$Hg, the shape configuration in the symmetric fusion path/valley
is two spherical shapes with $^{90}$Zr + $^{90}$Zr, which exists because
in the macroscopic model symmetric separated fragments are energetically
favored over asymmetric fragments, and the $N=50$ shell favors spherical
fragments.  The nascent fragment shapes in the symmetric fusion valley
for $^{198}$Hg are fairly deformed with $\epsilon=0.275$, because the
fragment neutron numbers are $N=59$, corresponding to onset of
deformation in separated nuclei.

An important feature for $^{180}$Hg is that the optimal potential-energy
curve from the ground state across the saddle and somewhat beyond and
the symmetric fusion valley are well separated by the potential ridge,
which initially is 8 MeV above the saddle region.  On the other hand,
the height of the corresponding ridge for $^{198}$Hg is much lower
(initially only 2 MeV high) and only persists for a narrow range in
$q_2$, suggesting that a change from the asymmetric shapes along the
initial fission path to different final fragment mass asymmetries is
less hindered in $^{198}$Hg than in $^{180}$Hg.

The separating ridge for $^{180}$Hg vanishes at $q_2=10.31$,
$\epsilon_{\rm H}=0.30$, $\epsilon_{\rm L}=0.15$, and $\alpha_{\rm
g}=0.20$ corresponding to $A_{\rm H}/A_{\rm L}=108.0/72.0$.  For
$^{198}$Hg, the ridge vanishes at $q_2=13.47$, $\epsilon_{\rm H}=0.40$,
$\epsilon_{\rm L}=0.0$, and $\alpha_{\rm g}=0.18$, corresponding to
$A_{\rm H}/A_{\rm L}=115.82/81.82$.  At the point where the separating
ridge vanishes, no ``obvious'' valley connects this location to a
scission configuration.  Instead we are on a rather flat
potential-energy surface which in the full 5D space gently slopes in
many directions. An analogy is being just below the top of a gently
sloping hill.  Therefore we cannot determine a plausible optimum fission
path by a static analysis alone.  However, when the neck is quite well
developed where the ridge disappears, it was suggested that the mass
asymmetry here might to a significant extent be preserved in the separated
fission fragments \cite{andreyev10:a}.
\begin{figure}[t]
\includegraphics[keepaspectratio,width=\linewidth]{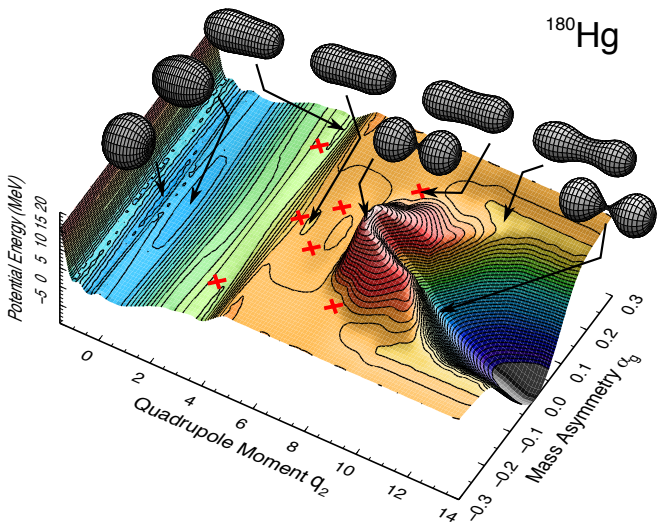}\\%
\caption{(Color online) Two-dimensional potential-energy surface for
$^{180}$Hg which shows some essential features of the full 5D
potential-energy surface.  Two crossed (red) lines show the location of some
saddle points.  Note in particular that the valley across the asymmetric
saddle disappears slightly beyond $q_2=10$} \label{Hg180-3D}
\end{figure}
\begin{figure}[t]
\includegraphics[keepaspectratio,width=\linewidth]{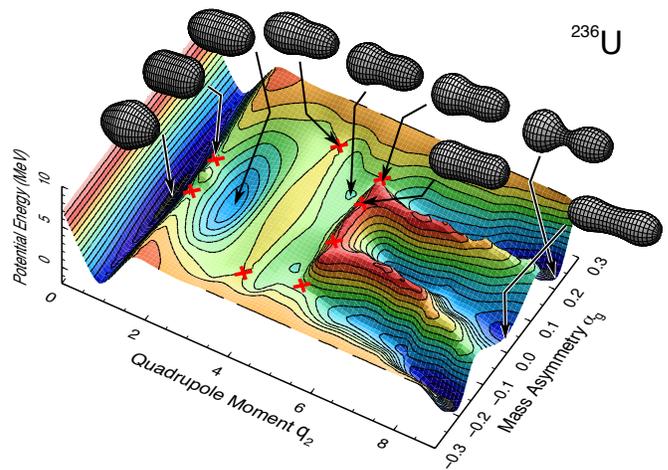}\\%
\caption{(Color online) Two-dimensional potential-energy surface for
$^{236}$U which shows some essential features of the full 5D
potential-energy surface.  Two crossed (red) lines show the location of some
saddle points.  Note in particular that the valley across the asymmetric
saddle continues to the largest $q_2$ shown. It also continues beyond to
a point where the nucleus separates into two fragments. This is very
much in contrast to the potential-energy surface for $^{180}$Hg.}
\label{U236-3D}
\end{figure}
\begin{figure*}[t]
\includegraphics[keepaspectratio,width=0.8\linewidth]{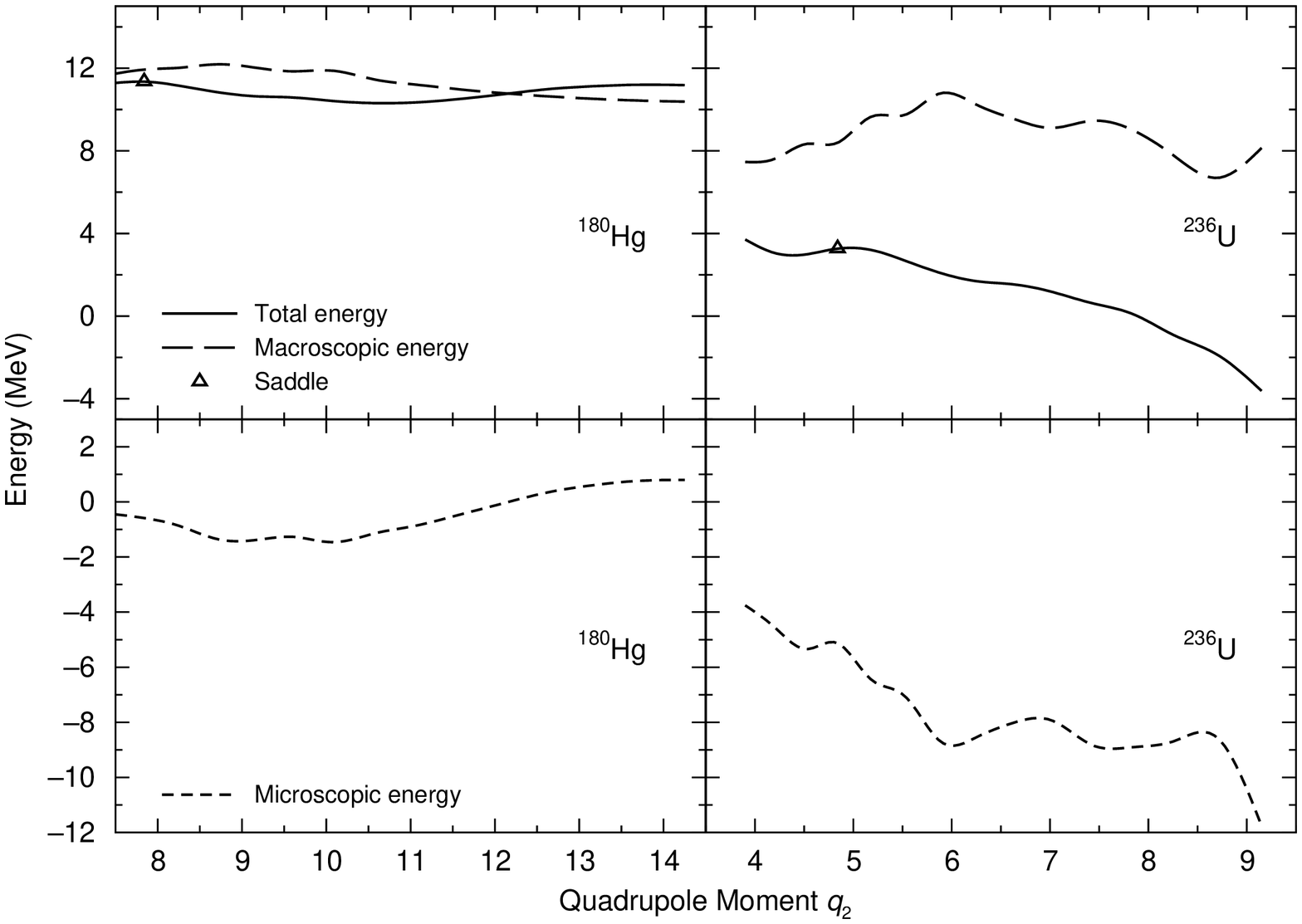}\\%
\caption{Total and macroscopic energies along the asymmetric fission
paths for $^{180}$Hg, and $^{236}$U are shown in the top two frames. The
microscopic energies along these paths are shown in the lower two
frames. There are very significant differences between the microscopic energy
 of the two nuclei.  } \label{sshellasympath}
\end{figure*}

\section{Saddle features and fission-fragment mass 
asymmetry in mercury isotopes} The fragment mass asymmetry in fission is
affected by the saddles, ridges and valleys in the fission
potential-energy surface that appear beyond the fission isomeric
minimum.  We have identified these features using the immersion method. The
results are summarized in Figs.~\ref{q2dep} and \ref{sadhgt}.  We pay
particular attention to the point where the ridge between the optimal
fission path and the fusion valleys disappears, which for $^{180}$Hg
occurs at $q_2=10.31$.  For the nuclei we study, it is not possible to
identify a clear mass-asymmetric fission path, because the ``fission
valley'' that takes us across the saddle point disappears at elongations
slightly beyond the saddle. That is, there is no continuous asymmetric
valley from the region of the saddle point to scission, very much in
contrast to the situation in the actinide region.

The mass-symmetric fusion path shown in Fig.~\ref{Hg180-1d} corresponds
to compact, nearly spherical fragment shapes. This type of fusion valley
is only present in Hg nuclei from $A=178$ to $A=190$. The ridges
separating the compact mass-symmetric fusion path become very low,
almost non-existent, at $A=190$, and this compact symmetric fusion
valley vanishes at $A=192$. Instead, for somewhat heavier isotopes a
mass-symmetric fusion path with large nascent-fragment deformations
appears.  To summarize, some general trends in the structure of the
potential-energy surfaces along the Hg isotope chain are:
\begin{itemize}
\item
With increasing $A$ the barrier height increases,
partly due to a lowering of the ground state as
$N=126$ is approached, and also to a decrease of fissility.
\item
The saddle shapes are more elongated (larger $q_2$)
for the heavier Hg isotopes.
\item
For low $A$ the ridges are prominent; 
for higher $A$, 
they almost disappear.
\end{itemize}

In the specific case of electron-capture-delayed fission of $^{180}$Hg
the shape asymmetry where the ridge vanishes could be related to the
observed fission-fragment mass asymmetry \cite{andreyev10:a}.  In
Fig.~\ref{wavedens} we show the asymmetry at this vanishing point for
the entire range of isotopes. We calculate the asymmetry from the
wave-function densities (top three panels), cf.\
Ref.~\cite{ichikawa09:a} for details.
In the bottom panel we show the
nascent-fragment shape-deformation parameters at this point.  These
features stand out:
\begin{enumerate}
 \item The  proton number of the light-(heavy)-mass fragment is close to $Z=34(46)$
       in all the Hg isotopes (see the second panel).
       However, no strong shell effect is present
       in the ground states of these fragments. The ground-state shapes
       for all the $Z=34(46)$ fragments are well deformed
       with uniformly positive microscopic corrections~\cite{moller95:b}.
 \item The neutron number of the light-mass fragment is close to $N=50$
       for $A > 190$ and the deformation 
       of those light fragments is spherical (cf. the third panel).
 \item At the vanishing point of the ridge
       the degree of fragment mass asymmetry becomes smaller with
       increasing mass number. But for the
       heavier isotopes the ridge is very short and low in energy so the
       asymmetry at the vanishing point might not be closely related to
       the final fragment mass asymmetry.
\end{enumerate}

\section{Two types of asymmetric fission}

Asymmetric fission in the actinide region has since its discovery been
``explained'' in terms of strong ``shells'' in the heavy fragment
related to its proximity to doubly magic $^{132}$Sn.  But it should be
observed that in fission of actinides the heavy fragment is not exactly
$^{132}$Sn and just small changes in $Z$ and $N$ from the doubly-magic
nucleus drastically decrease the extra binding due to proximity to a
doubly closed shell.  For example, the most probable heavy/light mass
split of $^{240}$Pu is $M_{\rm H}/M_{\rm L}= 140/100$. This corresponds
to the heavy fragment $^{140}_{\phantom{0}55}$Cs$_{\rm 85}$ with a
ground-state microscopic correction $-2.96$ MeV \cite{moller95:b}, which is
not even close to the $^{132}$Sn ground-state microscopic correction of
$-11.55$ MeV. But, when the nascent fragments start to emerge, they have
not absorbed some nucleons in the neck regions. Thus, the partially
formed heavy fragment in the case of $^{240}$Pu is closer in size and
shape to $^{132}$Sn than it is to $^{140}_{\phantom{0}55}$Cs$_{\rm 85}$,
which could significantly affect the microscopic correction. For example, just
removing one proton and one neutron from
$^{140}_{\phantom{0}55}$Cs$_{\rm 85}$ leads to
$^{138}_{\phantom{0}54}$Xe$_{\rm 84}$, with a ground-state microscopic
correction of $-5.35$ MeV \cite{moller95:b}.

Clearly, one should only invoke such hand-waving arguments related to
fragment properties as a starting point for understanding the
mass-asymmetric fission-fragment division in the actinide region. A more
complete understanding should involve the potential energy from the
ground-state shape to separated fragments in terms of a sufficiently
large number of shape degrees of freedom \cite{moller01:a}.  It has
indeed been shown that a deep asymmetric valley separated from a
symmetric fission valley for most actinides extends from the saddle
region to scission configurations \cite{moller01:a,moller09:a}. As an
example, we show in Fig.~\ref{U236-1d} calculated energies along symmetric and asymmetric
optimal fission paths and the separating ridge for $^{236}$U.  Here we
note an asymmetric valley extending from the outer saddle region to
scission. It is shielded from the symmetric valley by an about 5
MeV-high ridge along its entire path.  This contrasts very much with the
situation in the Hg region.

To illustrate more clearly the differences between Hg and actinides in
the fission potential-energy surfaces and the presence and absence of
``fragment'' shell effects in the potential-energy surfaces of the
compound system, we plot in Fig.~\ref{Hg180-3D} a by necessity somewhat
schematic two-dimensional representation of the most important features
of the full 5D potential-energy surface for $^{180}$Hg. In the left part
of Fig.~\ref{sshellasympath} we show the total energy, macroscopic
energy, and microscopic energy along a section of the asymmetric fission
path of $^{180}$Hg. In Fig.~\ref{U236-3D} and the right part of
\ref{sshellasympath} we show the corresponding quantities for $^{236}$U.

These figures illustrate visually the different origins of asymmetric
fission in the Hg and actinide regions. For $^{236}$U the asymmetric
valley extends from the outer saddle point to scission-like shapes. It
is a plausible assumption that the mean asymmetry in thermal
neutron-induced fission is close to the asymmetry of the shapes at the
bottom of the asymmetric valley. This correlation was indeed verified in
the investigation of Ref.~\cite{moller01:a} in which the calculated
asymmetry of the asymmetric valley bottom agreed with observed
fission-fragment mass asymmetries for 25 even-even actinide nuclides
with a mean deviation of only 3.0 nucleons.  The large negative microscopic
energy $E_{\rm sh} =-12$ MeV  at scission where $q_2=9$,
cf. Fig.~\ref{sshellasympath}, remains almost constant for more compact
shapes; it is still very substantial, $E_{\rm sh} =-6$ MeV at the
saddle-point deformation $q_2 = 5$.

In contrast, for $^{180}$Hg there is no valley extending from the saddle
region towards scission. Rather, for elongations only moderately beyond
the saddle the ridge separating the saddle region and the symmetric
fusion valley disappears. From static considerations alone it is not
obvious what trajectory towards separated fragments the nucleus will
follow.  Thus, as stated in Ref.~\cite{andreyev10:a}, the asymmetric
fission in $^{180}$Hg is a new type of asymmetric fission with its
origins in the local fission potential-energy-surface structure in the
saddle region, whereas in the actinide region a deep, persistent
asymmetric valley extends over the entire range from saddle-point shapes to
separated fragments. Figure \ref{sshellasympath} shows that there is no
significant fragment-related microscopic effect in the saddle region or
beyond for $^{180}$Hg; this microscopic energy is very low, fluctuating between $\pm 2$
MeV along the trajectory shown.

\section{Summary Discussion}

The recent observation of mass asymmetry in electron-capture delayed
fission of $^{180}$Hg \cite{andreyev10:a} has stimulated renewed
interest in fission since some simple ``fragment-shell'' type arguments
had anticipated that the most probable division would be into two
symmetric $^{90}$Zr fragments, because these exhibit two instances of
the spherical $N=50$ magic number and two instances of the spherical
$Z=40$ subshell.  It was proposed that a new type of asymmetric fission
had been observed, with its origins in the {\it local} structure in the
outer saddle-point region. Currently, the experimental data in this
neutron-deficient region in terms of energy range and number of nuclides
are extremely sparse, in particular in comparison with the data
available for heavier nuclei \cite{vandenbosch73:a,schmidt00:a} We have
calculated potential-energy surfaces of 12 even Hg isotopes in this
neutron-deficient region to establish the systematics of significant
structures. The most important finding is that it is only for nuclei in
the range $180\lnsim A \lnsim 190$ that the saddle region is somewhat
shielded from the symmetric fusion valley by a moderately high ridge
that also has some moderate extension in the elongation direction.  In
the $^{180}$Hg experiment the compound-nucleus excitation was limited to
about 1 MeV above the saddle point. This constraint and the ridge
structure allowed some qualitative conclusions about the expected
fragment asymmetries in this experiment \cite{andreyev10:a}.

In the actinide region numerous models have been proposed to describe
the observed fission mass asymmetries, for example
Refs.~\cite{fong56:a,wilkins76:a,brosa90:a,benlliure98:a,schmidt00:a,goutte05:a,randrup11:a}.
Often encouraging results are presented. We have shown here and
elsewhere \cite{moller01:a,moller09:a} that in calculated, realistic 5D
potential-energy surfaces, very strongly expressed, deep asymmetric
valleys are present.  These valleys usually
appear also in more approximate calculations, so that when the
respective model parameters are adjusted to experimental yields the
model results agree to varying degrees of accuracy with the experimental
data.  However none of these models have been applied to $^{180}$Hg,
with the exception of the Brownian shape-motion model
\cite{randrup11:a,moller12:a}, in which no parameter is adjusted.  Although the statistics of the
$^{180}$Hg experiment are limited, the Brownian shape-motion model may
be less accurate in this case than in the actinide region since for
$^{180}$Hg the result was $M_{\rm H}/M_{\rm L} = 104.4/75.6$ whereas the
experimental result was given as $M_{\rm H}/M_{\rm L} = 100/80$. More
striking is that the calculated FWHM width \cite{moller12:a} is about
twice the experimental result of 9 mass units~\cite{andreyev10:a}. In the actinide region the
calculated widths agreed very well with the experimental data
\cite{randrup11:a} with no obvious deviations except in the tails of the
yield distributions at very large asymmetries. A possible explanation of
these results is that in the actinide region the confining influence of
the steep walls of the asymmetric valley defines the width of the yield
distributions, and this feature is realistically described in the
calculations.  In the Hg region, where there are no confining ``fission
valley'' walls, the yield distribution is determined on the downslope of
a steep, smooth mountain side, cf. Fig.~\ref{Hg180-3D}. Here the fine
details of the dynamical part of the model may be more important than in
the actinide region. The models in the other
Refs. \cite{fong56:a,wilkins76:a,brosa90:a,benlliure98:a,schmidt00:a,goutte05:a}
have not yet been tested in this mass region.

Clearly, it will be a challenge to fission theories to reproduce
experimental data both in the Hg region and across the entire actinide
region without arbitrary model parametrizations which differ from region
to region.  Since we have now shown the different issues presented to
theory by fission in the Hg and actinide regions, we strongly encourage
efforts to obtain a more extensive set of fission data in the region
$180 \leq A \leq 200$ be undertaken, both in terms of excitation-energy
range and number of nuclides.  Such experiments would present new and
highly useful challenges to fission theories.

\begin{acknowledgments}
 A part of this research has been funded by MEXT HPCI
STRATEGIC PROGRAM.
 P.M. and A.J.S. acknowledge that this work was carried out under
 the auspices of the National Nuclear Security Administration of the US
 Department of Energy at Los Alamos National Laboratory under Contract
 DE-AC52-06NA25396, and the US Department of Energy through the LANL/LDRD
 Program. P.M. was also supported by a travel grant to JUSTIPEN
 (Japan-US Theory Institute for Physics with Exotic Nuclei) under Grant
 DE-FG02-06ER41407 (U. Tennessee).  The numerical calculations were
 carried out on SR16000 at YITP at Kyoto University.
\end{acknowledgments}                     


%

\end{document}